\documentclass[aps,prx,amsmath,nofootinbib,longbibliography,amssymb,superscriptaddress,twocolumn,10pt]{revtex4-1}

\usepackage[english]{babel}
\usepackage{graphicx}
\usepackage{ulem}
\usepackage{soul}
\usepackage{float} 
\usepackage[svgnames]{xcolor} 
\usepackage{comment}
\usepackage[colorlinks, linkcolor=NavyBlue , citecolor= NavyBlue,urlcolor = NavyBlue, breaklinks=true]{hyperref}
\usepackage{verbatim}
\usepackage{esint}
\usepackage{marginnote}

\renewcommand{\vec}[1]{\boldsymbol{#1}}
\newcommand{\RNum}[1]{\uppercase\expandafter{\romannumeral #1\relax}}

\def \bea {\begin{eqnarray}}
\def \eea {\end{eqnarray}}

\begin{document}

\title{Tunable non-Fermi liquid phase from coupling to two-level systems}

\author{Noga Bashan}\thanks{These authors contributed equally to this work.}
\affiliation{Department of Condensed Matter Physics, Weizmann Institute of Science, Rehovot 76100, Israel}

\author{Evyatar Tulipman}\thanks{These authors contributed equally to this work.}
\affiliation{Department of Condensed Matter Physics, Weizmann Institute of Science, Rehovot 76100, Israel}

\author{J\"{o}rg Schmalian}
\affiliation{Karlsruher Institut für Technologie, Institut f\"{u}r Theorie der Kondensierten Materie,  76049, Karlsruhe, Germany}

\affiliation{Karlsruher Institut für Technologie, Institut f\"{u}r Quantenmaterialien und Technologien,  76021, Karlsruhe, Germany}

\author{Erez Berg}
\affiliation{Department of Condensed Matter Physics, Weizmann Institute of Science, Rehovot 76100, Israel}

\date{\today}

\begin{abstract}
We study a controlled large-$N$ theory of electrons coupled to dynamical two-level systems (TLSs) via spatially-random interactions. 
Such a physical situation  
arises when electrons scatter off 
low-energy excitations in a metallic glass, such as a charge or stripe glass.
Our theory is governed by a non-Gaussian saddle point, which maps 
to the celebrated spin-boson model. 
By tuning the coupling strength we find that the model crosses over from a Fermi liquid at weak coupling to an extended region of non-Fermi liquid behavior at strong coupling, and realizes a marginal Fermi liquid at the crossover.
Beyond a critical coupling strength, the TLSs freeze and Fermi-liquid behavior is restored. Our results are valid for generic space dimensions $d>1$.
\end{abstract}

\maketitle

Understanding scenarios in which strong interactions between itinerant electrons and collective quantum fluctuations invalidate the conventional Landau Fermi liquid (FL) paradigm is a central problem in the field of correlated metals \cite{varma_colloquium_2020,chowdhury_sachdev-ye-kitaev_2022,hartnoll_planckian_2021,phillips_stranger_2022,sachdev_quantum_2023}. A prominent example of such a scenario is the enigmatic `strange metal' (SM) behavior found in high-$T_c$ superconductors and other quantum materials \cite{proust_remarkable_2019,bruin_similarity_2013,legros_universal_2019,cao_strange_2020}. SMs are commonly defined by an anomalous $T-$linear scaling of the dc resistivity at arbitrarily low temperatures, which is in sharp contrast to the $T^2$ dependence predicted by FL theory.
While SM behavior is often associated with an underlying quantum critical point (QCP) at $T=0$ \cite{gegenwart_quantum_2008}, there are multiple examples, such as cuprates \cite{cooper_anomalous_2009,hussey_tale_2018,hussey_generic_2013,greene_strange_2020} 
, twisted bilayer graphene \cite{cao_strange_2020,jaoui_quantum_2022}, and twisted transition metal dichalcogenides \cite{ghiotto_quantum_2021}, where 
it appears to persist over an extended region, suggesting the interesting possibility of a critical, non-Fermi liquid phase 
at zero temperature. 
In addition, extended critical behavior has been observed in MnSi with $T^{3/2}$ scaling of the resistivity \cite{pfleiderer_non-fermi_2003,pfleiderer_partial_2004,doiron-leyraud_fermi-liquid_2003,paschen_quantum_2021} and in CePdAl with a $T^n$ scaling with $n$ varying from $\sim1.4$ to $2$ \cite{zhao_quantum-critical_2019}.

The absence of a quasiparticle description in such systems, 
known as `non-Fermi liquids' (NFLs) \cite{anderson_new_1995,varma_singular_2002,lee_recent_2018} or `marginal Fermi liquids' (MFLs) \cite{varma_phenomenology_1989,varma_singular_2002}, 
makes well-controlled theoretical investigations 
challenging.
Nevertheless, recent years have witnessed a proliferation of illuminating solvable models, largely facilitated by the use of Large-$N$ and Sachdev-Ye-Kitaev (SYK) techniques \cite{sachdev_gapless_1993,Kitaev_SYK_talk,chowdhury_sachdev-ye-kitaev_2022,sachdev_quantum_2023}. In particular, the extensive analysis of a class of Yukawa-SYK theories \cite{marcus_new_2019, esterlis_cooper_2019,wang_quantum_2020,wang_solvable_2020,wang_phase_2021,esterlis_large_2021,guo_large_2022,valentinis_bcs_2023,valentinis_correlation_2023,patel_universal_2023}, where a Fermi surface is coupled to critical bosons, has demonstrated that strange metallicity can emanate from a quantum critical point. 

In contrast to strange metallicity that arises due to tuning to a QCP,
a controlled microscopic theory hosting a stable NFL phase, free of fine-tuning,  remains noticeably absent within the existing literature. This is surprising as  such extended metallic critical phases seem to be allowed within  holographic setups \cite{zaanen_planckian_2019,hartnoll_horizons_2011}, and are also supported by numerical evidence \cite{wu_non-fermi_2022}.

\begin{figure}[t]
\centering
 \includegraphics[width=1\columnwidth]{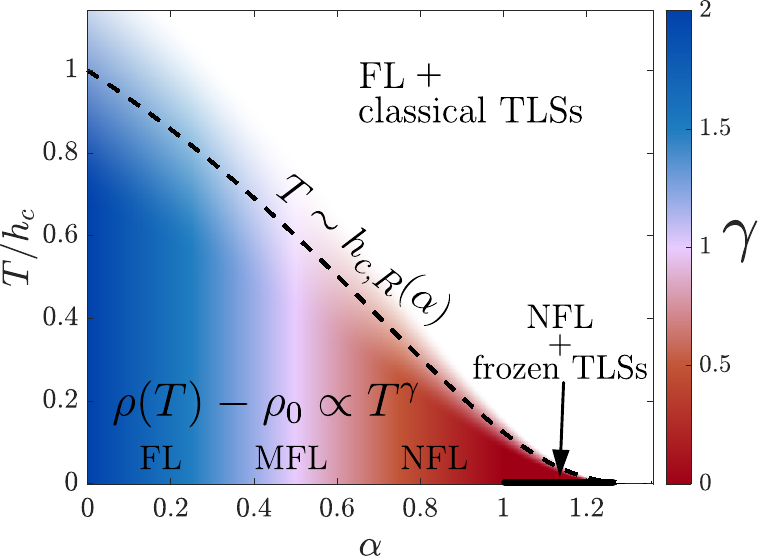}
\caption{Phase diagram for a system  with a linear TLS distribution ($\beta=1$) and $h_c/E_F=0.4$. $\alpha$ is the dimensionless coupling strength and $h_c \ll E_F$  is the bare cutoff energy of the TLS distribution.  The colored area represents the low-$T$ dynamical regime, characterized for $\alpha<1$ by the exponent $\gamma = 2(1-\alpha)$ that governs the  inelastic scattering rate. We find
FL behavior for $\alpha<1/2$, a MFL at $\alpha=1/2$ and a NFL for $\alpha>1/2$. As $\alpha$ crosses $1$, the TLSs partially freeze at $T=0$, indicated by a thick black line. At temperatures above the renormalized cut off $T>h_{c,R}$, the TLSs act as classical elastic scatterers with $T-$independent scattering rate (plus $T^2/E_F$ corrections). The  crossover to this region is denoted by the fading out of the color.
}
\label{fig:phase diagram of x model}
\end{figure}

In this paper we show that a NFL phase can arise due to coupling of electrons to the low-energy two-level excitations of an underlying glassy order. 
To this end we formulate and solve in a controlled large-$N$ limit a model of $N$ fermion species coupled to $M\sim N$ dynamical two-level systems (TLSs) per site via spatially random interactions. The theory is
governed by a non-Gaussian saddle point, which maps to the spin-boson (SB) model.
While the electrons constitute an ohmic bath for the TLSs for all coupling strengths, 
the back action of the TLSs gives rise to a tunable exponent $\gamma(\alpha)$ in the electronic self energy  ${\rm Im}\Sigma (\omega )\sim \left| \omega \right |^\gamma$. Here $\alpha$ is the dimensionless coupling constant of the problem; see below.
 Notably, the low-temperature dc resistivity obeys $\rho(T) =\rho_0 + CT^{\gamma}$, with a tunable exponent $0<\gamma<2$ that depends on the coupling.
 This behavior offers an alternative viewpoint to the conventional form: $\rho(T) = \rho_0 + AT + BT^2$, that is often used in the interpretation of strange metal behavior \cite{park_isotropic_2008,cooper_anomalous_2009,doiron-leyraud_correlation_2009,hussey_dichotomy_2011,jin_link_2011,hussey_tale_2018,legros_universal_2019,berben_compartmentalizing_2022}. 
Our theory draws inspiration from the intricate phase diagrams of high$-T_c$ superconductors which often exhibit a competition between multiple frustrated orders that could lead to metallic glassiness \cite{emery_frustrated_1993,kivelson_stripe_1998}. It is worth noting that such a scenario can arise independently of extrinsic quenched disorder, namely, due to frustration-induced self-generated randomness \cite{schmalian_stripe_2000}. 

We start from the following model of $N$ species of electrons hopping on a $d-$dimensional hypercubic lattice with dispersion $\varepsilon_{\boldsymbol{k}}$. Each site hosts $M$ two-level systems (i.e. spin$-1/2$ Pauli operators) subject to random fields $h_{l,\boldsymbol{r}}$ which we locally describe in their eigenbases:
\begin{equation}
    H = \sum_{\boldsymbol{k},i<N} \varepsilon_{\boldsymbol{k}}  c_{i\boldsymbol{k}}^{\dagger}c_{i\boldsymbol{k}}-\sum_{\boldsymbol{r},l<M}h_{l,\boldsymbol{r}} \sigma^z_{l,\boldsymbol{r}} + H_{\rm int}.
\end{equation}
Having multiple TLSs per site ($M\gg1$) could arise from a coarse grained description of mesoscale collective degrees of freedom of an underlying glassy state~\cite{lubchenko_intrinsic_2001,lubchenko_origin_2003}, each of which interacts with many electrons.  
The coupling between the electrons and TLSs is given by
\begin{equation}
    H_{\rm int} = \frac{1}{N}\sum_{\boldsymbol{r},ijl} \boldsymbol{g}_{ijl,\boldsymbol{r}}\cdot \boldsymbol{\sigma}_{l,\boldsymbol{r}} c_{i\boldsymbol{r}}^{\dagger}c_{j\boldsymbol{r}}.
    \label{Hamiltonian}
\end{equation}
Here, the coupling constants are also assumed to be random numbers. It is natural to add to this problem a term with potential scattering $H_{\rm pot} = \frac{1}{\sqrt{N}}\sum_{\boldsymbol{r},ij<N}  V_{ij,\boldsymbol{r}} c_{i\boldsymbol{r}}^{\dagger}c_{j\boldsymbol{r}}$ that induces elastic scattering 
of the itinerant electrons, while leaving the TLS properties unaffected.

Clearly, for $\boldsymbol{g}_{ijl,\boldsymbol{r}}\sim (0,0,1)$ the TLSs are static and the problem reduces to a single-particle one and can be solved exactly. Electrons cause quantum transitions whenever there are coupling constants that differ from the above choice. We have analyzed the case of general coupling constants and find that the case where $\boldsymbol{g}_{ijl,\boldsymbol{r}}\sim (g_{ijl,\boldsymbol{r}},0,0)$ describes all the important  aspects that occur in the generic case \cite{longTLS}. 
Finally, we need to specify the distribution function of the random fields and coupling constants. For the latter we chose a Gaussian distribution with zero mean and variance $\overline{g_{ijl,\boldsymbol{r}
} g_{ijl,\boldsymbol{r}'
}}=g^2\delta(\boldsymbol{r}
-\boldsymbol{r}
')$.
The low-energy collective excitations in a glass are effectively described by a collection of two-level systems (TLS)~\cite{anderson_anomalous_1972,phillips_tunneling_1972}.  Each TLS represents 
two nearly-degenerate configurations of some local collective mode. The level splitting and tunneling rate, characterizing each TLS, can be considered as independently distributed random variables due to their disordered nature. For the eigenvalues $h$ in Eq.~\eqref{Hamiltonian} this yields a distribution function ${\cal P}(h)\propto |h|$ for $|h|<h_c$, where $h_c$ is the maximum splitting of the TLSs, assumed to be much smaller than the Fermi energy, $E_F$. In principle, we can consider a more general behavior ${\cal P}_\beta(h)\propto |h|^\beta$ that also includes the case of a constant distribution for $\beta=0$ (which could describe a physical situation where tunneling is significantly smaller than the energy splitting). We will keep $\beta\geq0$ arbitrary. The qualitative behavior is, however, similar for all values of $\beta\geq 0$.

In order to solve the problem we employ the formalism developed in the context of Yukawa-SYK models~\cite{marcus_new_2019, esterlis_cooper_2019,wang_quantum_2020,wang_solvable_2020,wang_phase_2021,esterlis_large_2021,guo_large_2022,valentinis_bcs_2023,valentinis_correlation_2023,patel_universal_2023}. We start from a coherent-state path integral for the fermions and spin (i.e. TLS) variables, average over the random couplings by using the replica trick, and enforce the identities 
\begin{eqnarray}
    G_{\vec{r},\vec{r}'}(\tau,\tau') = \frac{1}{N}\sum_i \overline{c}_{i\vec{r}}(\tau)c_{i\vec{r}'}(\tau') , \label{Gfermion} \\
    \chi_{\vec{r}}(\tau,\tau') = \frac{1}{M}\sum_l \sigma_{l,\vec{r}}^{x}(\tau)\sigma_{l,\vec{r}}^{x}(\tau'), \label{Dtls} 
\end{eqnarray}
via corresponding Lagrange multiplier fields $\Sigma_{\vec{r},\vec{r}'}(\tau,\tau')$ and $\Pi_{\vec{r}}(\tau,\tau')$. 
One can integrate out the fermions and obtains the effective action:
 \begin{eqnarray}
     S_{\rm eff} & = & S_{\rm TLS}-N{\rm tr}\log\left(G_{0}^{-1}-\Sigma\right)-N{\rm tr}G\Sigma+\frac{M}{2}{\rm tr}\chi\Pi \nonumber \\
     &+&\frac{M}{2}g^{2}\sum_{\boldsymbol{r}}\int_{\tau,\tau'} G_{\boldsymbol{r}}\left(\tau,\tau'\right)G_{\boldsymbol{r}}\left(\tau',\tau\right)\chi_{\boldsymbol{r}}\left(\tau,\tau'\right),
 \end{eqnarray}
where the trace is taken over space and time. In contrast to the usual Yukawa-SYK approach, where the role of the TLSs is played by Gaussian bosonic modes, there is no Wick theorem for spin variables and the TLSs cannot be integrated out. Instead, they are governed by the action $S_{\rm TLS}=\sum_{\boldsymbol{r},l} S_{\rm SB}\left[\boldsymbol{\sigma}_{l,\boldsymbol{r}}\right]$ which is a sum of individual spins with fermion-induced dynamics:
\begin{eqnarray}
    S_{\rm SB}\left[\boldsymbol{\sigma}_{l,\boldsymbol{r}}\right]&=&-\int_{\tau,\tau'} \Pi_{\boldsymbol{r}}\left(\tau'-\tau\right)\sigma_{l,\boldsymbol{r}}^{x}\left(\tau\right)\sigma_{l,\boldsymbol{r}}^{x}\left(\tau'\right)\nonumber \\
    &-&\int_{\tau} h_{l,\boldsymbol{r}}\sigma_{l,\boldsymbol{r}}^{z}(\tau)+S_{\rm Berry}\left[\boldsymbol{\sigma}_{l,\boldsymbol{r}}\right].
    \label{eq:spin_bos}
\end{eqnarray}
 See Supplementary Information (SI) for details of the derivation of Eq. \eqref{eq:spin_bos}. The last term is merely the Berry phase that occurs from the spin coherent state representation. $S_{\rm SB}\left[\boldsymbol{\sigma}_{l,\boldsymbol{r}}\right]$ is a highly nontrivial local many-body problem. One recognizes easily that it is equivalent to the celebrated spin-boson problem of a localized spin coupled to an environment of bosons with spectral function that leads to the bath function  $\Pi_{\boldsymbol{r}}\left(\tau'-\tau\right)$~\cite{caldeira_quantum_1983,leggett_dynamics_1987,costi_thermodynamics_1999,weiss_quantum_2012}.  In our case the bath is 
 due to electronic particle-hole excitations. 
In the large-$N,M$ limit the theory is governed by the saddle point with respect to 
to $G$, $\Sigma$, and $\chi$, given respectively by
\begin{eqnarray}
\Sigma_{\boldsymbol{r},\boldsymbol{r}'}\left(\tau\right)&=&\delta_{\boldsymbol{r},\boldsymbol{r}'}\frac{M}{N}g^{2}G_{\boldsymbol{r},\boldsymbol{r}}\left(\tau\right)\chi_{\boldsymbol{r}}\left(\tau\right)  \nonumber \\ 
G_{\boldsymbol{r},\boldsymbol{r}'}\left(i\omega\right)&=&\left.\left(G_{0}^{-1}\left(i\omega\right)-\Sigma\left(i\omega\right)\right)^{-1}\right|_{\boldsymbol{r},\boldsymbol{r}'} \nonumber \\  
\Pi_{\boldsymbol{r}}\left(\tau\right) & = & -g^{2}G_{\boldsymbol{r},\boldsymbol{r}}\left(\tau\right)G_{\boldsymbol{r},\boldsymbol{r}}\left(-\tau\right).
\label{eq:saddle}
\end{eqnarray}
The saddle with respect to $\Pi$ replaces the right-hand-side of Eq.~\eqref{Dtls} by its expectation value. Eqs.~\eqref{eq:saddle} are similar to the Yukawa-SYK problem, where one obtains essentially a set of coupled Eliashberg equations. The crucial difference is that in our case, $\Pi$ and $\chi$ are not the self energy and propagator of a boson, related by a Dyson equation. Instead $\Pi$ is the bath function of a spin-boson problem whose solution determines $\chi$.

The saddle-point equations \eqref{eq:saddle} together with the solution of the spin-boson problem $S_{\rm SB}$ of \eqref{eq:spin_bos} are valid for a given disorder configuration $\left\{ h_{l,\boldsymbol{r}}\right\}$ of the random field. Hence, the system is not translation invariant and correlation functions like $\left\langle \sigma_{l,\boldsymbol{r}}^{x}\left(\tau\right)\sigma_{l,\boldsymbol{r}}^{x}\left(0\right)\right\rangle $ fluctuate in space. However, 
to determine the self energy in Eq.~\eqref{eq:saddle} we only  need to know the average of this correlation function over the $M$ flavors. 
To proceed we employ the large-$M$ limit 
to replace sums over the TLS flavors with an average over the TLS splitting distribution: $\frac{1}{M}\sum_{l=1}^{M} \cdots \to \int \mathcal{P}\left(h\right) \cdots d h$. This substitution does not necessitate self-averaging in the replica sense, but rather follows directly from the central limit theorem. 
 Since the  distribution is independent of position we obtain a statistical translation invariance for the average  of interest. Hence, $\chi(\tau)$ and therefore the bath function $\Pi(\tau)$ are independent of $\boldsymbol{r}$, implying that  the local fermionic Green's function and self energy are  position-independent as well.

The theory is therefore governed by a momentum-independent self-energy. Standard manipulations for such a self energy, valid in the regime where the typical bosonic energies are smaller than the electronic bandwidth, yield $\Pi(i\omega) = \frac{\rho_F ^2 g^2 }{2\pi}|\omega|$. That is, the electrons constitute an ohmic bath for the TLSs, independent of the form of the self energy. To proceed, we obtain the TLS-susceptibility
$\chi(h_l,\omega)$ (i.e. for a fixed $h_l$) by solving its corresponding 
SB problem $S_{\rm SB}\left[\boldsymbol{\sigma}_{l,\boldsymbol{r}}\right]$, and average over the random field configurations $\chi(\omega)=\int dh {\cal P}\left(h\right) \chi(h,\omega)$. Then the fermionic self-energy of Eq.~\eqref{eq:saddle} can at $T=0$ be written as ${\rm Im}\Sigma_{\rm ret}\left(\omega\right)=-\rho_F g^2(M/N)\int_{0}^{\left|\omega\right|}d\omega'{\rm Im}\chi_{\rm ret}\left(\omega'\right)$ (here and henceforth `ret' denotes the retarded function). Extensions to finite temperatures are straightforward.

\begin{figure}[t]
\centering
\includegraphics[width=1\columnwidth]{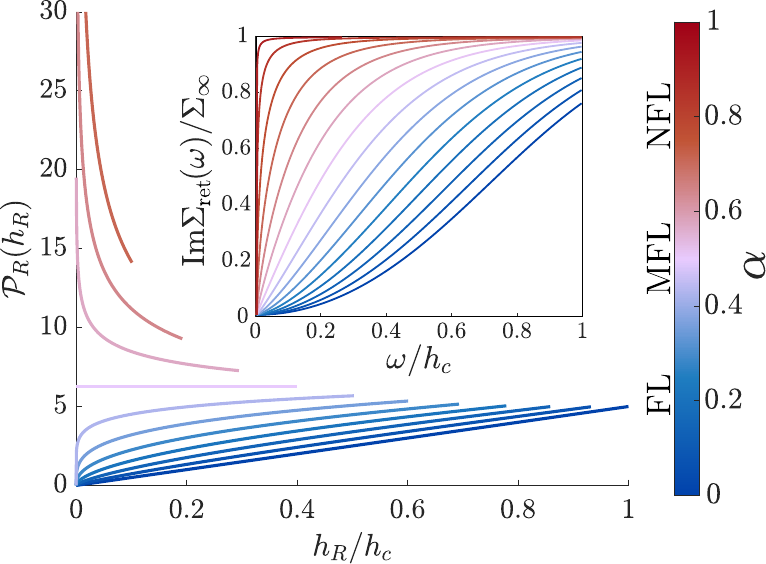}
\caption{Main: Evolution of the renormalized distribution $\mathcal{P}_R(h_R)$ (see Eq.~\eqref{eq:renormalized dist}) is shown for increasing dimensionless coupling $\alpha$ with $\beta=1$ and $h_c/E_F=0.4$. 
Inset: Frequency dependence of ${\rm Im}\Sigma_{\rm ret}(\omega)$ at $T=0$ for different $\alpha$. ${\rm Im}\Sigma_{\rm ret}(\omega\ll h_{c,R})\sim |\omega|^{2(1-\alpha)}$,  while it saturates to the value $\Sigma_\infty\sim \alpha E_F$  for $\omega \gg h_{c,R}$ due to the finite Hilbert space of TLSs.
}
\label{fig:evolution of distribution}
\end{figure}

The solution of the SB model is a classic problem in many-body physics of impurity models \cite{weiss_quantum_2012,leggett_dynamics_1987}.
For our purposes we use two established facts: (i) the low-energy physics of the problem is governed by renormalization group equations
\begin{equation}
  \frac{d\alpha}{d\ell} = -\alpha\tilde{h}^2\,\,\, {\rm and }\,\,\, \frac{d\tilde{h}}{d\ell} = (1-\alpha) \tilde{h}  
  \label{eq:floweq}
\end{equation}
for the dimensionless coupling constant 
$\alpha=g^2 \rho_F^2 / \pi^2$ and field $\tilde{h}= h/E_F$, where $\ell$ measures the logarithm of the characteristic energy. Its solution yields renormalized fields and coupling constants $h_R(h,\alpha)$ and  $\alpha_R(h,\alpha)$, respectively (up to $\mathcal{O}(1)$ numerical coefficients which can be determined by alternative methods \cite{hur_entanglement_2008,camacho_exact_2019,filyov_method_1980,ponomarenko_resonant_1993}). (ii) The correlation function obeys a scaling form in terms of the renormalized field
\begin{eqnarray}
    {\rm Im}\chi_{\rm ret}(h,\omega,T)=\frac{1}{\omega}f_\alpha\left(\frac{\omega}{h_R},\frac{T}{h_R}\right).
    \label{eq:scalchi}
\end{eqnarray}
The scaling function $f_\alpha$ is known numerically and, in several limiting regimes, analytically. To perform the average over field configurations, it is more convenient to work with the distribution function of renormalized fields:
\begin{equation}
    {\cal P}_{R,\beta} \left(h_R\right)\equiv\frac{dh}{dh_R}{\cal P}_\beta\left(h\right).
\label{eq:renormalized dist}
\end{equation}
The renormalized distribution function follows from the solution of the flow equations. In Fig.~\ref{fig:evolution of distribution} we show the evolution of the corresponding renormalized distribution function as function of coupling constant $\alpha$ for the case of $\beta=1$ (i.e. a linear bare distribution).

We begin our analysis in the regime $\alpha<1$. In this case the solution of Eqs.~\eqref{eq:floweq} yields $h_R= E_{F}\tilde{h}^{\frac{1}{1-\alpha}}$\cite{caldeira_quantum_1983,leggett_dynamics_1987,costi_thermodynamics_1999,weiss_quantum_2012}, which yields that ${\cal P}_{R,\beta} \left(h_R\right)\propto h_R^{\beta-\alpha\left(1+\beta\right)}$.
The downwards renormalization of the excitation energies of the TLSs leads to a strong weight transfer in the distribution function, making it significantly more singular at low energies. As soon as $\alpha$ reaches $\alpha_{\rm MFL}\equiv\beta/(1+\beta)$ the renormalized distribution function remains finite at arbitrarily small $h_R$, for $\alpha>\alpha_{\rm MFL}$ it is even divergent in the low-energy limit. We can now straightforwardly perform the average with the renormalized distribution. 
We find that at $T=0$, the low-energy ($\omega \ll h_{c,R}$) TLS-susceptibility and the electronic self-energy are characterized by a coupling-constant-dependent exponent $\gamma=(1+\beta)(1-\alpha)$ (see SI for details):
\begin{eqnarray}
    {\rm Im}\chi_{\rm ret}\left(\omega\right) &=& {\rm sign}(\omega)   \frac{\gamma A_\alpha}{h_{c,R}}\left|\frac{\omega}{h_{c,R}}\right|^{\gamma-1},\nonumber\\
    {\rm Im}\Sigma_{\rm ret}\left(\omega\right)&=&- \rho_F g^2 A_\alpha \frac{M}{N} \left|\frac{\omega}{h_{c,R}}\right|^{\gamma}.
\end{eqnarray}
Here, $ h_{c,R}$ is the renormalized value of the upper cut-off $h_c$. Importantly, the exact form of the scaling function $f_\alpha(x,0)$ of Eq.~\eqref{eq:scalchi} only affects the $\mathcal{O}(1)$ coefficient $A_\alpha$, given in the SI.  At finite temperatures  Eq.~\eqref{eq:scalchi} implies that the averaged TLS-correlation function and hence the self energy obey $\omega/T$-scaling. Since the exponent $\gamma$ is a continuous function of the coupling strength $\alpha$, the self-energy can be tuned to realize a marginal Fermi-liquid form: ${\rm Im}\Sigma_{\rm ret}\left(\omega\right)\sim |\omega|$, provided that $\alpha = \alpha_{\rm MFL}$. The MFL behavior arises at generic couplings and is not associated with a quantum critical point \cite{sachdev_quantum_2011}, rather, the system smoothly crosses over between a Fermi liquid for $\alpha<\alpha_{\rm MFL}$, where the lifetime is large compared to the excitation energy, and a non-Fermi liquid for $\alpha>\alpha_{\rm MFL}$.

\begin{figure}[t]
\centering

\includegraphics[width=1\columnwidth]{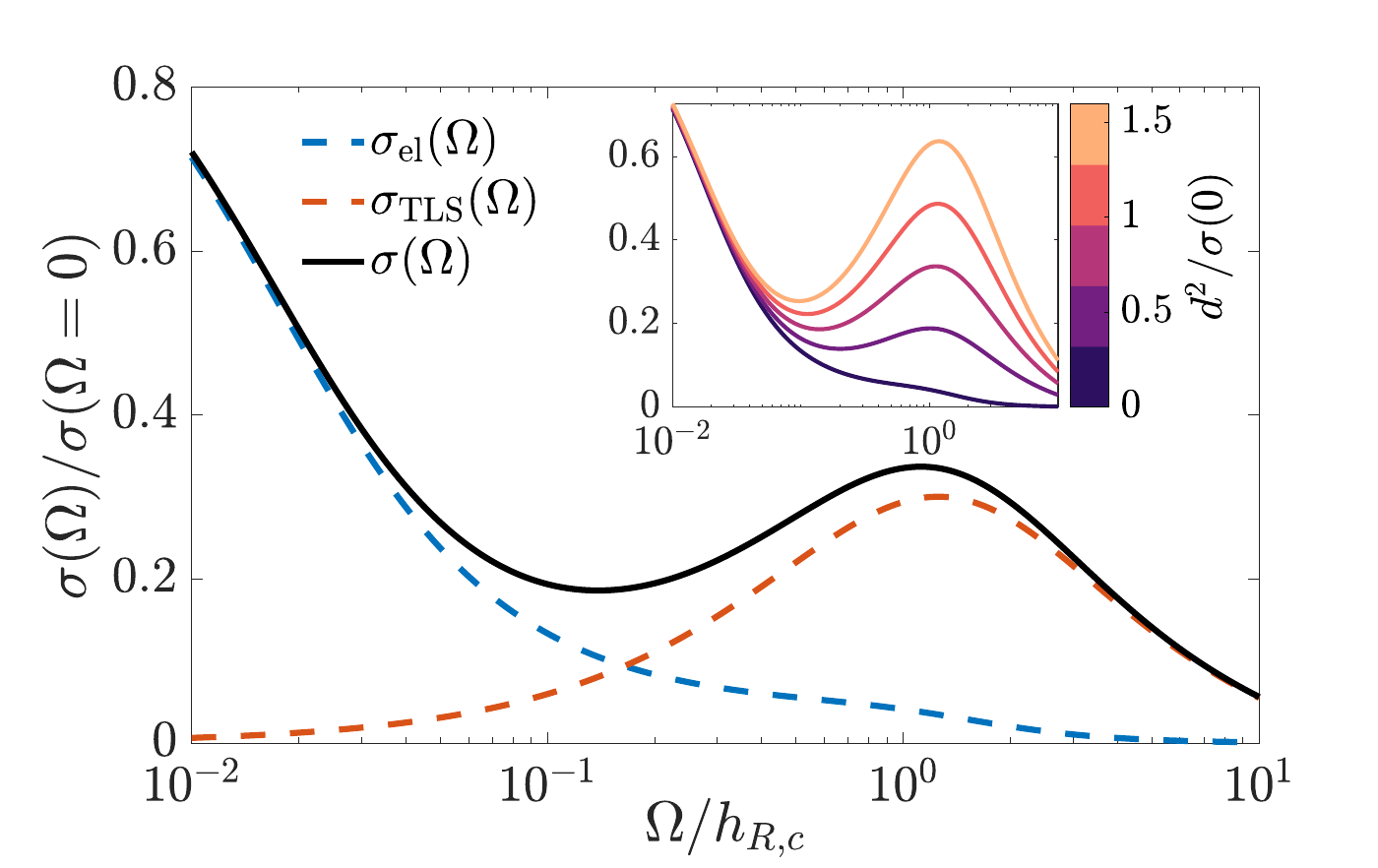}
\caption{Representative plot of the optical conductivity at the MFL point ($\gamma = 1$). The electronic and TLS contributions are shown in dashed lines. Inset shows the total optical conductivity for different values of the TLS-dipole moment strength $d^2$.}
\label{fig:optical cond}
\end{figure}

We proceed to consider transport properties of the theory using the Kubo formula. With vanishing vertex corrections due to spatially uncorrelated interactions, 
the $T$-dependence of the conductivity follows from the frequency dependence of the self energy. We find 
\begin{equation}
    \rho(T\ll  h_{R,c})-\rho_0 \propto T^\gamma,
\end{equation}
 where the residual resistivity, $\rho_0$, is due to the potential scattering term $H_{\rm pot}$. The same reasoning for the thermal conductivity implies that the Wiedemann-Franz law is obeyed in the $T\rightarrow 0$ limit \cite{tulipman_criterion_2022}.
For $T\gtrsim h_{R,c}$, $\rho(T)$ saturates to a $T-$independent constant, up to a small $T^2/E_F$ correction. Our results agree with the MFL behavior and corresponding $T$-linear resistivity previously obtained in the weak coupling limit with $\beta=0$ and $\alpha\to 0$ \cite{black_resistivity_1979}. The optical conductivity is given by $\sigma(\Omega) = \sigma_{\rm el}(\Omega) + \sigma_{\rm TLS}(\Omega)$, where the first term denotes the electronic contribution and the second term is due to dipole excitations of TLSs, with magnitude proportional to the typical TLS-dipole moment. At $T=0$, in the absence of potential scatterers we find $\sigma_{\rm el} \propto \Omega^{-{\rm min}(2-\gamma,\gamma)}$ ($\gamma=1$ admits a logarithmic correction $\sim\Omega^{-1}\log^{-2}(\Omega)$), while the TLS term reads $\sigma_{\rm TLS}(\Omega) \propto \overline{d^2} \Omega {\rm Im}\chi_{\rm ret}(\Omega)$, $\overline{d^2}$ being the typical squared magnitude of the TLS-dipole moment, and hence may
give rise to a non-monotonic $\Omega$-dependence of the optical response. The $T$-dependence of the resistivity and the frequency dependence of $\sigma(\Omega)$, with small potential scattering term included, are shown in Fig.~\ref{fig:optical cond}. 

\allowdisplaybreaks{
A similar analysis can be performed for a coupling constant $\alpha=1$ and $\alpha>1$. For $\alpha=1$ the renormalized field is exponentially small, $h_{R}\sim E_{F}\exp\left(-\tfrac{\pi}{2\tilde{h}}\right)$, 
which yields a renormalized distribution function ${\cal P}_{R}\left(h_{R}\right)\sim h_R^{-1}\log^{-(2+\beta)}\left(\frac{E_{F}}{h_{R}}\right)$
and we find for the TLS propagator and fermionic self energy the highly
singular behavior
\begin{eqnarray}
{\rm Im}\chi_{\rm ret}\left(\omega\right) & \sim & \frac{1}{\omega\log^{2+\beta}\left(\frac{E_{F}}{\left|\omega\right|}\right)},\nonumber \\
{\rm Im}\Sigma_{\rm ret}\left(\omega\right) & \sim & \frac{1}{\log^{1+\beta}\left(\frac{E_{F}}{\left|\omega\right|}\right)},
\label{eq:alphaeq1}
\end{eqnarray}
where the single-particle scattering rate vanishes extremely slowly
as $\omega\rightarrow0$, a behavior owed to the fact that there are
many exponentially small excitations of the ensemble of TLSs (this is reminiciesnt of the behavior found in \cite{giamarchi_singular_1993}). 

For $\alpha>1+h/E_F$ the flow equations imply that the field $\tilde{h}$ flows to zero
and the TLS freezes due to the Caldeira-Leggett mechanism \cite{caldeira_influence_1981,Chakravarty1982,Chakravarty1984,leggett_dynamics_1987,weiss_quantum_2012}. This yields
a contribution $\delta\left(h_{R}\right)$ to the renormalized distribution
function and a corresponding term $\sim \delta\left(\omega\right)$ in
${\rm Im}\chi\left(\omega\right)$, giving rise to a constant elastic scattering term in ${\rm Im}\Sigma\left(\omega\right)$.
The frozen TLSs  behave like additional potential scatterers
and give an extra contribution to the residual resistivity, i.e. Fermi
liquid behavior is reinstated. In the narrow regime $1<\alpha<1+h_{c}/E_{F}$  we must divide the TLSs into those which are still dynamical ($h/E_F>\alpha-1$), and those which are frozen ($h/E_F\leq\alpha-1$). While the frozen ones give rise to a constant  ${\rm Im}\Sigma_{\rm ret}\left(\omega\right)$, the contribution of the dynamical ones will be similar to that  for $\alpha=1$ given in Eq.~\eqref{eq:alphaeq1}.

In conclusion, we formulated and solved a model for two-level systems  coupled to electrons that are expected to emerge in a glassy metallic state. As a result we find a critical phase with exponents that change as one varies non-thermal parameters that influence the dimensionless coupling constant, i.e. that affect the electronic density of states or the interaction matrix element. Electrons form an ohmic bath for the TLS, while the back action of TLSs yields the variable exponent for electrons. We then find a 
sequence of crossovers from a Fermi liquid via a marginal Fermi liquid and to a non-Fermi liquid state as one increases the coupling strength; once the interaction becomes too strong, all TLSs freeze and one recovers Fermi liquid behavior. We verified that these results are robust if we include more generic coupling constants $\boldsymbol{g}_{ijl,\boldsymbol{r}}$, even though this requires introducing multiple ohmic baths, provided that the dominant bath is not aligned with the direction of the field \cite{longTLS} (i.e. that the $z$ component of $\vec{g}^2$ is not the largest of the three).

Our model is solvable in the large-$N$,$M$ limit. 
Physically, this limit may describe the case where the TLSs are spatially extended, and interact with many channels of incident itinerant electrons, each with a random interaction (corresponding to the many electron `flavors' in our model). Nevertheless, it is important to mention how $1/N$ corrections may modify the results at small energies. 
One effect neglected in our analysis is RKKY-like interactions between 
TLSs (mediated by itinerant electrons), which are suppressed by $1/N$ due to the frustrated nature of the interactions.  Including this 
effect,
we find that each TLS modifies 
the bath felt by other TLSs \cite{dobrosavljevic_absence_2005,tanaskovic_spin-liquid_2005}, and its ohmic behavior
breaks down below a small energy scale $\omega_{\star}\sim E_F/N^{\frac{1}{2-\gamma}}$. Another interesting $1/N$ effect concerns the validity of the self-averaging assertion, which we find to hold only above energies of order $\omega_{\star \star}\sim h_{c,R}/N^{\frac{1}{\gamma}}$. We defer further
investigation of such effects to future study \cite{longTLS}.

Our analysis can be extended to investigate the role of TLS fluctuations for the formation of Cooper pairs. We expect that they will cause or strengthen existing pairing interactions that are particularly pronounced at intermediate coupling.  Our theory is applicable in arbitrary space dimensions $d>1$, i.e. the quantum critical phase is not due to strong long-wave fluctuations that are particularly pronounced in low dimensions, but due to the very slow dynamics present in an ensemble of local degrees of freedom. It is conceivable that such slow local fluctuation of a glassy state are responsible for some of the strange metal behavior observed in strongly correlated materials.
}

\textit{Acknowledgements.---}
This work was motivated and inspired by prior discussions and unpublished work with G. Grissonnanche, S. A. Kivelson, C. Murthy, A. Pandey, B. Ramshaw, and B. Spivak on the physics of two-level systems in electronic glasses and their possible role in strange metals.
We also thank N. Andrei, A. V. Chubukov, V. Dobrosavljevic, T. Holder, P. A. Lee, Y. Oreg, S. Sachdev, A. Shnirman and C. Varma for helpful discussions. 
J.S. was supported by the German Research Foundation (DFG) through CRC TRR 288 ``ElastoQMat,'' project B01 and a Weston Visiting Professorship at the Weizmann Institute of Science. 
E.B. was supported by the European Research Council (ERC) under grant HQMAT (Grant Agreement No. 817799) and by the Israel-US Binational Science Foundation (BSF). 
Some of this work was performed at KITP, supported in part by the National Science Foundation under PHY-1748958.
\bibliography{TLSs.bib}

\appendix
\onecolumngrid

\section{Mapping to the spin-boson model}

In this Appendix, we provide further details on the mapping of our model to the spin-boson problem with an ohmic bath (Equations 6, 7 in the main text). We set $V^2=0$ for simplicity, as its inclusion makes no difference in the derivation below. We specialize to the case where $\vec{h}\propto \hat{z}, \vec{g}\propto \hat{x}$, although the derivation of the most general case is essentially the same and will be shown in \cite{longTLS}.

We begin by considering the spin coherent-state path integral representation for the TLSs. The partition function for a given configuration of $h$ and $g$ is given by 
\begin{eqnarray}
    Z[h,g]= \int \mathcal{D}[\boldsymbol{\sigma},c,\overline{c}] e^{-S},
\end{eqnarray}
with the action $S=S_0 + S_{\rm int}$:
\begin{eqnarray}
    S_0 &=& \sum_{\boldsymbol{r}}\sum_{l=1}^{M}S_{\rm Berry}[\boldsymbol{\sigma}_{l,\boldsymbol{r}}] - \sum_{\boldsymbol{r}}\sum_{l=1}^{M} \int_{\tau} h_{l,\boldsymbol{r}} {\sigma}^z_{l,\boldsymbol{r}} \nonumber\\&+& \sum_{i=1}^{N}\sum_{\boldsymbol{k}}\int_{\tau} \overline{c}_{i\boldsymbol{k}}(\partial_\tau +\varepsilon_{\boldsymbol{k}}) c_{i\boldsymbol{k}},  \\
    S_{\rm int} &=& \frac{1}{N} \sum_{\boldsymbol{r}}\sum_{i,j=1}^{N}\sum_{l=1}^{M}\int_{\tau} {g}_{ijl,\boldsymbol{r}}{\sigma^x}_{l,\boldsymbol{r}}  \overline{c}_{i\boldsymbol{r}}c_{j\boldsymbol{r}}.
\end{eqnarray}
Here we kept the same symbols $\boldsymbol{\sigma}_{l,\boldsymbol{r}}$ for the unit vectors that result from the coherent state representation of the Pauli operators. $S_{\rm Berry}$ denotes the Berry's phase of the TLSs, see e.g. \cite{auerbach_interacting_1998}. 

To proceed, we average over the random couplings $g_{ijl,\vec{r}}$ using the replica method, and introduce the bilocal fields in Eqs.~\eqref{Gfermion} and \eqref{Dtls}, which we enforce
via conjugated fields, $\Sigma$ and $\Pi$, respectively\footnote{Notice that, for simplicity, we are considering spinless fermions. In this case, there is no pairing instability to leading order in $1/N$. In a model of spinful fermions, one would have to consider also the anomalous part of the Green's function, and an instability towards a superconducting state with an intra-flavor, on-site order parameter may occur~\cite{esterlis_cooper_2019}.}.

Next, we integrate over the fermions and substitute a replica-diagonal \textit{Ansatz}, which allows us to express the partition function as  
$Z[h]=\int\mathcal{D}\left[G,{\chi},\Sigma,\Pi,\vec{\sigma} \right]e^{-S_{\rm eff}}$, where the effective action is given by

\begin{eqnarray}
   S_{\rm eff}&=&-N\text{Tr}\ln\left(G_{0}^{-1}-\Sigma\right)-N\int_{\tau,\tau'}\sum_{\boldsymbol{r},\boldsymbol{r}'}\sum_{\sigma}G_{\boldsymbol{r},\boldsymbol{r}'}\left(\tau,\tau'\right)\Sigma_{\boldsymbol{r},\boldsymbol{r}'}\left(\tau,\tau'\right)+\frac{M}{2}\int_{\tau,\tau'}\sum_{\boldsymbol{r}}\chi_{\boldsymbol{r}}\left(\tau,\tau'\right)\Pi_{\boldsymbol{r}}\left(\tau',\tau\right)\nonumber\\&+&\frac{M}{2}\int_{\tau,\tau'}\sum_{\boldsymbol{r}}g^{2}G_{\boldsymbol{r}}\left(\tau,\tau'\right)G_{\boldsymbol{r}}\left(\tau',\tau\right)\chi_{\boldsymbol{r}}\left(\tau,\tau'\right)+\sum_{\boldsymbol{r}}\sum_{l=1}^{M}S_{\text{Berry}}\left[\boldsymbol{\sigma}_{l,\boldsymbol{r}}\right]-\int_{\tau}\sum_{\boldsymbol{r}}\sum_{l=1}^{M}{h}_{l,\boldsymbol{r}} {\sigma^z}_{l,\boldsymbol{r}}\nonumber\\&-&\frac{1}{2}\int_{\tau,\tau'}\sum_{\boldsymbol{r}}\Pi_{\boldsymbol{r}}\left(\tau',\tau\right)\sum_{l=1}^{M}\sigma^x_{l,\boldsymbol{r}}\left(\tau\right)\sigma^x_{l,\boldsymbol{r}}\left(\tau'\right).
\end{eqnarray}
The limit of large $M$ and $N$, with fixed ratio $M/N$, justifies the use of the saddle point approximation. Performing the variation with respect
to $G$ and $\Sigma$ gives the first two lines in Eqs.~\eqref{eq:saddle},
where we have used thermal equilibrium to write the saddle-point equations  with time-translation-invariant  correlation functions and their Fourier transforms.
In addition, the stationary point that follows from the variation
with respect to $\chi$ is
\begin{eqnarray}
\Pi_{\boldsymbol{r}}\left(\tau\right) & = & -g^{2}G_{\boldsymbol{r},\boldsymbol{r}}\left(\tau\right)G_{\boldsymbol{r},\boldsymbol{r}}\left(-\tau\right).
\label{eq:saddlePi}
\end{eqnarray}
The Berry phase term $S_{\rm Berry}$, that reflects the fact that no Wick theorem exists for Pauli operators, implies that  the TLSs cannot be simply integrated over as a Gaussian integral. However,  it allows us to recast the TLS problem to that of $M$ decoupled TLSs per site $\boldsymbol{r}$, 
$\sum_{\boldsymbol{r},l=1}^{M}S_{\boldsymbol{r},l}\left[\boldsymbol{\sigma}_{\boldsymbol{r},l}\right]$, coupled to a bosonic bath of particle-hole excitations.
Each TLS is governed by the spatially local effective action
\begin{eqnarray}  S_{\boldsymbol{r},l}\left[\boldsymbol{\sigma}\right]=S_{{\rm Berry}}\left[\boldsymbol{\sigma}\right]-\int_{\tau}{h}_{l,\boldsymbol{r}}{\sigma^z}\left(\tau\right)-\int_{\tau,\tau'}\Pi_{\boldsymbol{r}}\left(\tau'-\tau\right)\sigma^{x}\left(\tau\right)\sigma^{x}\left(\tau'\right).
\label{effectiveTLSaction}
\end{eqnarray}
 This is indeed the action of the spin-boson model after the bosonic bath degrees of freedom have been integrated out~\cite{weiss_quantum_2012} (Eq. 6 in the main text). The latter give rise to the non-local in time coupling $\Pi_{\boldsymbol{r}}^{a}\left(\tau',\tau\right)$ that is, in general, different for each site.  Of course, in our problem the  origin of the bath function are not bosons but the conduction electrons. For the solution of this local problem this makes, however, no difference. 
 $S_{\boldsymbol{r},l}$ still depends on the random configuration ${h}_{l,\boldsymbol{r}}$ of the fields.

For a given realization of the fields ${h}_{l,\boldsymbol{r}}$
the problem is not translation invariant and correlation functions like $\left\langle \sigma_{l,\boldsymbol{r}}^{x}\left(\tau\right)\sigma_{l,\boldsymbol{r}}^{x}\left(0\right)\right\rangle $ fluctuate in space. However, 
to determine the self energy in the second line of \eqref{eq:saddle} we only need to know the average $\chi_{\vec{r}}(\tau)$ of this correlation function over the $M$ TLS-flavors. 
To proceed, we employ the central limit theorem, considering the correlation function $\chi_{\vec{r}}(\tau)$ as a random variable, we may replace the sum of the
TLS flavors 
with averaging over the TLS splitting distribution ($\sum_{l=1}^{M} \to M\int \mathcal{P}\left({h}_{\boldsymbol{r}}\right) d{h}_{\boldsymbol{r}}$). Since the distribution $\mathcal{P}$ is independent of position, the self-averaging assumption translates to a statistical translation invariance of the model, at least for the average  of interest. Hence, $\chi_{\boldsymbol{r}}(\tau)=\chi(\tau)$ is independent on $\boldsymbol{r}$. The same must then hold for the bath function $\Pi_{\boldsymbol{r}}(\tau)=\Pi(\tau)$. From the  saddle point equations \eqref{eq:saddlePi} it follows that  the local fermionic Green's function (and through Eqs.~\eqref{eq:saddle}, the self energy) are both space independent. 
Hence we can go to momentum space and find that
the theory is goverened by a momentum-independent self-energy and the Dyson equation for the electrons read
\begin{eqnarray}  
\label{Dyson}
\Sigma\left(\tau\right)&=&\frac{M}{N}g^{2}\chi\left(\tau\right)G\left(\tau\right),\\
G_{\boldsymbol{k}}\left(i\omega\right)&=&\frac{1}{i\omega-\varepsilon_{\boldsymbol{k}}-\Sigma\left(i\omega\right)},
\end{eqnarray}
where $G\left(\tau\right)=\int_{\boldsymbol{k}}G_{\boldsymbol{k}}\left(\tau\right)$ is the local Green's function. 
For a momentum independent fermionic self energy we obtain in the limit of large electron bandwidth
\begin{eqnarray}
G\left(i\omega\right)=\int_{\boldsymbol{k}}G_{\boldsymbol{k}}\left(i\omega\right)\approx-i\pi\nu_{0}{\rm sgn}\left(\omega\right).
\label{GF_av}
\end{eqnarray}
The particle-hole correlation function can now be evaluated. We find
\begin{equation}
    \Pi(\omega) = \frac{\nu_0^2 g^2 }{2\pi}|\omega|,
\end{equation} 
irrespective of the electronic self-energy. We thus conclude that each TLS is coupled to an ohmic bath of particle-hole excitations that is independent of the back reaction of the TLS on the electronic degrees of freedom. Thus, we have shown that the (spatially local) TLS-correlator
\begin{eqnarray}
    \chi\left(\tau-\tau'\right)&=&\frac{1}{M}\sum_{l}\left\langle \sigma_{l}^{x}\left(\tau\right)\sigma_{l}^{x}\left(\tau'\right)\right\rangle, 
    \label{DysonTLS}
\end{eqnarray}
is determined by the behavior of $M$ decoupled SB models. 

The strategy of the solution of our model in the large-$N$ limit is therefore: 
(i) solve the spin-boson problem  with ohmic bath for a given realization of the field $\boldsymbol{h}$, (ii) average over the TLS distribution function of the fields, and (iii) use the resulting propagator $\chi_a(\omega)$ of the TLSs to determine the fermionic self energy from Eq.\eqref{Dyson}. The non-linear character of the problem is rooted in the rich physics of the spin-boson problem, along with the averaging over the distribution functions of the fields ${h}$.
The dimensionless coupling parameter of the relevant SB model is related to the interaction strengths by:
\begin{equation}
    \alpha = \frac{\nu_0^2g^2}{\pi^2}.
\end{equation}

\section{Derivation of Equations 11, 13}

At $T=0$, the average (imaginary part of the) susceptibility is given by
\begin{eqnarray}
    \overline{\text{Im}\chi_{\rm ret}} &=& \int_0^{h_c} \text{Im}\chi_{\rm ret}(\omega,h) \mathcal{P}_\beta(h)dh  \\
    &=& \int_0^{h_{R,c}} \frac{1}{\omega} f_\alpha \left(\frac{\omega}{ h_R},0\right) \mathcal{P}_R( h_R) d h_R \\
    &=& {\rm{sgn}}(\omega) \int_{|\omega|/h_{R,c}}^\infty \frac{f_\alpha \left(x,0\right)}{x^2} \mathcal{P}_R(|\omega|/x) dx 
\end{eqnarray}
The result of this integral thus depends on the renormalized distribution $\mathcal{P}_r$.
\subsection{$\alpha<1$}
Starting with $ h_R=c_\alpha \omega_c (h/\omega_c)^{1/(1-\alpha)}\Rightarrow h = ( h_R/c_\alpha)^{1-\alpha}\omega_c^{-\alpha}$:
\begin{eqnarray}
    \mathcal{P}( h_R) &=& \mathcal{P}(h) \left(\frac{dh}{d h_R}\right)\\
    &=& \mathcal{N}  h_R^{\beta(1-\alpha)-\alpha}
\end{eqnarray}
Since the distribution is cut off at $h_{R,c}= h_R(h_c)$, the normalization constant must be $\mathcal{N}=\gamma/h_{R,c}^\gamma$, with $\gamma=\beta(1-\alpha)-\alpha+1=(1+\beta)(1-\alpha)$. Inserting this into the averaged susceptibility gives:
\begin{eqnarray}
    \overline{\text{Im}\chi_{\rm ret}} &=& {\rm{sgn}}(\omega) \int_{|\omega|/h_{R,c}}^\infty \frac{f_\alpha \left(x,0\right)}{x^2} \frac{\gamma|\omega|^{\gamma-1}}{h_{R,c}^\gamma x^{\gamma-1}} dx \\
    & =& \frac{1}{\omega} \left|\frac{\omega}{h_{R,c}}\right|^\gamma \times \gamma \int_{|\omega|/h_{R,c}}^\infty \frac{f_\alpha \left(x,0\right)}{x^{\gamma+1}} dx
    \label{eq:chiforAalpha}
\end{eqnarray}
Using that fact that at long times $\chi(t) \propto \alpha/t^2$, we see that for small frequencies $\text{Im}\chi_{\rm ret}(\omega\ll h_R)\propto \alpha\omega \Rightarrow f_\alpha(x\ll1)\propto \alpha x^2$. Near the lower integration limit the integrand is $\propto 1/x^{\gamma-3}$. If $\gamma<2$ then the integral converges when taking $\omega/h_{R,c}\to0$, and can thus be considered as a constant of order unity, and we find the frequency dependence $\sim\omega^{\gamma-1}$ as presented in the main text. (If $\gamma>2$ then the integral diverges as $|\omega|^{2-\gamma}$ and the resulting overall frequency dependence is $\overline{\text{Im}\chi_{\rm ret}}\propto \omega$. This is because the averaged susceptibility cannot decay faster than the susceptibility of the TLSs with highest $h$.)
Considering Eq.~\eqref{eq:chiforAalpha}, we observe that 
\begin{eqnarray}
    A_\alpha=\int_{0}^\infty \frac{f_\alpha(x,0)}{ x^{\gamma+1}}dx
\end{eqnarray} 
The scaling function $f_\alpha$ is not known analytically for generic values of $\alpha$, but, as previously mentioned, is supported mainly around $x\sim 1$ such that the integral is $\mathcal{O}(1)$. For example at the Toulouse point $A_{1/2} = 2$ for a linear distribution. 

\subsection{$\alpha\approx1$}
\label{Appendix: alpha=1}
In this case we use $ h_R=c_\alpha\omega_c\exp(-b\omega_c/h)
\Rightarrow h=\frac{b\omega_{c}}{2\log\left(\frac{c_\alpha \omega_{c}}{ h_R}\right)}$, which gives the renormalized distribution:
\begin{eqnarray}
\label{P Delta_r alpha=1}
    \mathcal{P}( h_R)
    &=& \frac{\mathcal{N}}{ h_R\log^{2+\beta}\left(\frac{c_\alpha\omega_c}{ h_R}\right)}
\end{eqnarray}
and the normalization can be found to be $\mathcal{N}=(1+\beta)\log^{1+\beta}(\omega_c/h_{R,c})$. We neglect for simplicity the factor of $c_\alpha \sim \mathcal{O}(1)$ inside the logarithm. The averaged susceptibility is then given by
\begin{eqnarray}
    \overline{\text{Im}\chi_{\rm ret}} &=& \frac{1+\beta}{\omega}\log^{1+\beta}(\omega_c/h_{R,c}) \int_{|\omega|/h_{R,c}}^\infty \frac{f_\alpha \left(x\right)}{x \log^{2+\beta}\left(x\omega_c/|\omega|\right)} dx 
\end{eqnarray}
In order to simplify the integral, we rely on the fact that $f_\alpha(x\gg1)\propto 1/x^{4-2\alpha}$ \cite{guinea_dynamics_1985} and $f(x\ll1)\propto x^2$, such that most of the weight of the integral is around $x\sim\mathcal{O}(1)$, for which $|\log(x)|\ll \log(\omega_c/|\omega|)$. Thus we may neglect the $x$ dependence inside the log and take it out of the integral. We thus retrieve the frequency dependence presented in the main text, and the the $x$ integreal, as before, only affects the numerical prefactor.

\end{document}